\documentclass[authorversion,nonacm]{acmart}

\usepackage{hhline}
\usepackage{xcolor}
\usepackage{multirow}
\usepackage{colortbl}

\newcommand\blfootnote[1]{
    \begingroup
    \renewcommand\thefootnote{}\footnote{#1}
    \addtocounter{footnote}{-1}
    \endgroup
}

\AtBeginDocument{%
  }

\copyrightyear{2025}
\acmYear{2025}
\setcopyright{cc}
\setcctype{by}

\renewcommand\footnotetextcopyrightpermission[1]{}

\begin{document}

\title[The Gulf of Interpretation: From Chart to Message and Back Again]{The Gulf of Interpretation: From~Chart~to~Message~and~Back~Again}

\author{Christian Knoll}
\email{christian.knoll@univie.ac.at}
\orcid{0000-0002-6560-4273}
\affiliation{
  \institution{Faculty of Computer Science, Doctoral School Computer Science, University of Vienna}
  \city{Vienna}
  \country{Austria}
}

\author{Torsten Möller}
\email{torsten.moeller@univie.ac.at}
\orcid{0000-0003-1192-0710}
\affiliation{
  \institution{Faculty of Computer Science, Data Science @ Uni Vienna, University of Vienna}
  \city{Vienna}
  \country{Austria}
}

\author{Kathleen Gregory}
\email{k.m.gregory@cwts.leidenuniv.nl}
\orcid{0000-0001-5475-8632}
\affiliation{
  \institution{Centre for Science and Technology Studies, Leiden University}
  \city{Leiden}
  \country{Netherlands}
}

\author{Laura Koesten}
\email{laura.koesten@univie.ac.at}
\orcid{0000-0003-4110-1759}
\affiliation{
  \institution{Faculty of Computer Science, University of Vienna}
  \city{Vienna}
  \country{Austria}
}

\renewcommand{\shortauthors}{}

\begin{abstract}
Charts are used to communicate data visually, but often, we do not know whether a chart's intended message aligns with the message readers perceive. In this mixed-methods study, we investigate how data journalists encode data and how members of a broad audience engage with, experience, and understand these visualizations. We conducted workshops and interviews with school and university students, job seekers, designers, and senior citizens to collect perceived messages and feedback on eight real-world charts. We analyzed these messages and compared them to the intended message. Our results help to understand the gulf that can exist between messages (that producers encode) and viewer interpretations. In particular, we find that consumers are often overwhelmed with the amount of data provided and are easily confused with terms that are not well known. Chart producers tend to follow strong conventions on how to visually encode particular information that might not always benefit consumers.
\end{abstract}



\maketitle

\blfootnote{\copyright 2025 Copyright held by the owner/authors. This is the authors' version of the work. It is posted here for your personal use. Not for redistribution. The definitive Version of Record will be published in CHI Conference on Human Factors in Computing Systems (CHI '25), April 26-May 1, 2025, Yokohama, Japan. \url{http://doi.org/10.1145/3706598.3713413}.}

\section{Motivation}
\label{sec:introduction}
Charts are critical to communicate data to the general public~\cite{engebretsen2020}. This is true in times of crisis (e.g., the latest COVID-19 pandemic, natural disasters, war) but also in communicating data that has an impact on the daily lives of individuals (e.g., elections, inflation, job market). Despite this, 
there is extensive work in the visualization community on effective visual encodings for the perception of data~\cite{cleveland1984}. Further, there are established standards for evaluating visual tools for domain experts through frameworks such as The Nested Model~\cite{munzner2009} and the Design Study Methodology~\cite{sedlmair2012}. However, we know less about the cognitive and affective aspects of understanding charts meant to communicate particular messages to a broad set of people~\cite{lan2024}. Thus, we argue that a deeper understanding of these aspects is needed to bridge the \textit{gulf of interpretation} between what a producer intends and what a reader understands. Creating effective visualizations that can be understood and made sense of by a broad set of people (e.g., the general public) is challenging and, when done poorly, might lead to confusion, misunderstanding, or even distrust~\cite{franconeri2021}.

What information does a consumer take away from a visualization? How do received messages differ across diverse consumer groups (e.g., young students, senior citizens) and by chart type? How do received messages match the intended message of the visualization producer?

To tackle these questions, we designed a mixed-methods study to explore how visualization producers encode data and how visualization consumers engage with and understand these data visualizations. In the first step of our methodology, we collaborated with visualization producers who earn their living by designing charts for broad audiences, such as data journalists and people working with them. These producers provided us with charts that they created for a public audience. Through a series of workshops in the form of focus groups as well as interviews with diverse audiences, we studied what consumers take away from a visualization---a visualization's \textit{message}---and how they would rate a visualization on different subjective measures (e.g., understandability, readability, trustworthiness). The resulting messages were then analyzed and returned to the producers, whom we asked to re-design their charts in light of the consumers' interpretation and experience. These re-designed charts were then compared to the original charts to see whether they matched, providing us insights into possible gaps between producer intent and consumer interpretation.

The analysis revealed several interesting findings regarding chart interpretation by diverse user groups. Despite being categorized as an ``easy'' chart by the producer, the stacked bar chart was the hardest to understand in all aspects (harder than a Sankey diagram). Further, people stumbled over unconventional chart designs or not well-clarified terms and were quickly overwhelmed by the amount of data to comprehend. To our surprise, collaborative sensemaking activities did not add anything to the messages identified by individuals. We also found that our professional chart designers often refer to conventional representations when designing charts based on messages (instead of data). For instance, they chose Sankey diagrams to compare voting behaviors in two elections, despite being aware that they are not easily understood, and line charts for time-varying data. However, in contrast to consumer workshops, here the collaborative setting encouraged novel ideas. These also divert from convention, abstracting away from showing the complete data to readers towards a more direct encoding of the envisioned message.

This work enabled us to gather in-depth insights into conveying messages about data through visualizations and their interpretation through diverse audiences from multiple perspectives. The contributions of this work include:

\begin{itemize}
    \item a novel approach to study the \textit{gulf of interpretation} between chart producers and consumers of messages conveyed by (data) visualization,
    \item in-depth insights into the interpretation of charts from a diverse participant sample who are not commonly part of visualization evaluation research and
    \item we add to and build on insights into the impacts of visualization design choices regarding different chart types, including aesthetic or emotional aspects that impact chart engagement.
\end{itemize}

\section{Related Work}
\label{sec:relatedwork}

In this section, we review the literature on different target audiences of visualizations, the sensemaking of visualizations through the articulation of takeaway messages, and approaches used to evaluate visualizations in group settings, which informed the development of the method used in this work.

\textbf{Understanding target audiences of visualizations.}
Given the vast design space of visual data representations, designing effective visualizations commonly includes efforts to create a clear picture of the target audience(s), including their tasks and requirements. This is also known as \textit{collaborative visualization research}, where visualization researchers work closely with domain experts to develop visual analysis tools~\cite{akbaba2023}. Frameworks such as The Nested Model~\cite{munzner2009} and the Design Study Methodology~\cite{sedlmair2012} focus on domain experts for whom visual analysis systems are designed. However, visualizations are often used in media to communicate information to a broad set of people (e.g., the general public), which cannot be defined as a single, isolated, homogeneous target audience~\cite{lee2020,jena2021}. There is limited research on how different audiences interpret charts. Recently, Franconeri et al.~\cite{franconeri2021} reviewed guidelines for effective and intuitive visualizations oriented toward different audiences, including students, coworkers, and the general public. Participant samples in visualization research tend to have limited diversity~\cite{solen2022}, but there are some notable examples of efforts to study non-experts. There is still a discrepancy in the visualization community on a definition for such non-experts~\cite{burns2023}. For instance, a study by Price et al.~\cite{price2016} shows how visualization can facilitate older adults' decision-making when choosing a drug prescription plan. He et al.~\cite{he2023} investigate how general audiences engage with visualizations of public data and found varying levels of audiences' openness to receive new information. Peck et al.~\cite{peck2019} conducted interviews in rural Pennsylvania to explore how usually excluded populations in visualization studies perceive visualizations. They found that different factors (e.g., personal experience, chart simplicity, chart aesthetics) impact how people prioritize visualizations and that ``[...] data visualization studies performed primarily with highly educated students may not generalize to the larger public [...]''~\cite[p. 10]{peck2019}, which informed our participant recruitment.

\textbf{Studying data visualization literacy through messages and the \textit{gulf of interpretation}.}
Data visualization literacy---the ability to read and create data visualizations---is becoming increasingly important~\cite{boerner2019,pandey2023}. There is literature on how to measure visualization literacy through assessment tests, e.g., using Item Response Theory~\cite{boy2014}, VLAT~\cite{lee2017}, Mini-VLAT~\cite{pandey2023}, or CALVI~\cite{ge2023}. Another way of measuring how people understand visualizations~\cite{lee2016,burns2020,karer2020} is by looking at what messages~\cite{koesten2023,stokes2023,vaidya2020} or insights~\cite{north2006,chang2009} they take away from them, an approach we adopt in our methodology. Measuring understanding is part of the sensemaking process of visualizations, which is still poorly understood~\cite{pohl2017}. Studies have also looked at sensemaking in collaborative settings~\cite{mahyar2014}, in contrast to individual sensemaking, arguing that ``it is important to create a shared understanding of the information available to achieve shared goals''~\cite[p. 322]{paul2010}. 

Related to Norman's term ``gulf of execution''~\cite{norman1998}, which refers to the difference between the user's intention and what a system allows them to do, we are interested in investigating how diverse people interpret charts and how this compares and contrasts to the chart's intended message. We assume that a \textit{gulf of interpretation} exists between what a producer intends and what a reader understands. This is not least due to the diversity of factors influencing peoples' understanding of data visualizations, some of which are more subjective than others~\cite{kennedy2016}. Vaidya and Dasgupta~\cite{vaidya2020} address the \textit{gulf of interpretation}, which they refer to as \textit{communication gap}, by contributing a fact-evidence reasoning framework for decoding facts in visualizations. While this theoretical framework is aimed at helping chart producers and consumers, our work is based on feedback and opinions from actual chart producers and consumers. Dasgupta et al. also conducted an exploratory study to bridge a gap between theory (visualization researchers) and practice (domain experts) by analyzing ``how climate scientists use and design visualizations for reflecting upon the causes and effects of design problems''~\cite[p. 996]{dasgupta2015}. Hullman and Diakopoulos~\cite{hullman2011} show that framing visualizations through certain design techniques can influence the interpretation of a chart. This is also reflected by Lee-Robbins and Adar~\cite{leerobbins2023}, who argue that the affective intents of chart designers can influence the chart reader's opinions, attitudes, or values. Additionally, we want to analyze and compare the content of readers' messages to identify which levels of information are considered important from a visualization. Lundgard and Satyanarayan~\cite{lundgard2022} define four levels of semantic content that can be used to classify such messages. Alternatively, Koesten et al.~\cite{koesten2023} focus on four categories (granularity, composition, structure, and content) for characterizing messages. In this work, we use both approaches to characterize messages since they complement each other.

\textbf{Development of the methodological approach.}
Tory~\cite{tory2014} reflects on different types of user studies and categorizes them based on their goals and methodological approaches. Placing our work in this categorization, we aimed to understand and evaluate data visualizations with different audiences through quantitative and qualitative approaches. A substantial amount of literature focuses on the evaluation of visualizations~\cite{lam2012,tory2005,zuk2006} using different methods, including experimental studies, focus groups (workshops), surveys, and interviews~\cite{rohrer2022}. Creative Visualization-Opportunities (CVO) workshops~\cite{kerzner2019,knoll2020} are used for the requirement analysis of visual analysis tools in close collaboration with domain experts. Inspired by this framework, which aims to create visual analysis solutions for a domain problem through a mix of workshop activities (usually starting from data and arriving at visualization designs), we used a reversed process starting at the visualization to see how chart readers make sense of it. We decided to use workshops as one of our methods as they can produce richer results through collaborative participant activities than individual interviews~\cite{massey1991}. Additionally, they allowed us to combine activities that facilitate individual and collaborative sensemaking (of our readers) and chart creation (of our data journalists). However, we faced recruitment issues with senior citizens for the group setting, a challenge also highlighted by Witschey et al.~\cite{witschey2013}, which is why we opted for individual interviews with this group. Finally, our approach was informed by Groen and Polst~\cite{groen2021} on conducting interviews tailored to senior citizens.

\begin{figure*}[t]
\includegraphics[width=\textwidth]{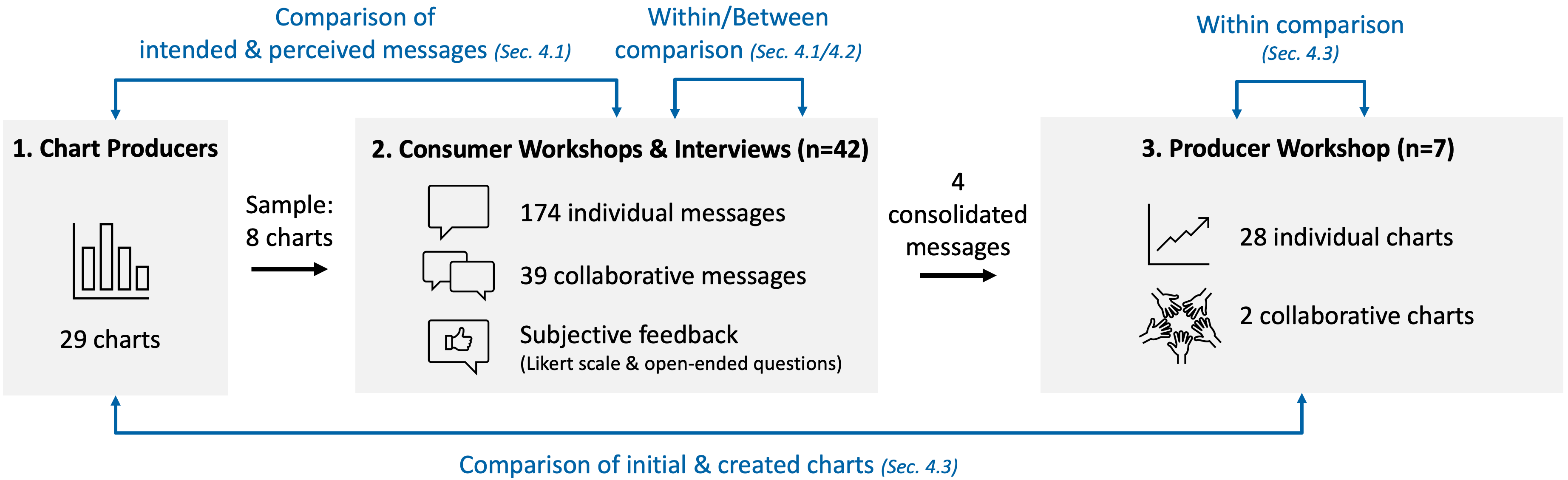}
\caption{Our methodological process with three phases. The blue arrow lines depict the different parts of the analysis and the corresponding section.}
\label{fig:method_process}
\Description[Figure 1 shows a research study workflow involving chart producers and consumers.]{Figure 1 shows a research study workflow involving chart producers and consumers. The process starts with 29 charts created by chart producers, from which eight are selected. These charts are then evaluated by 42 participants in workshops and interviews, generating individual and collaborative messages and feedback. The results are condensed into four messages, which are used in a workshop with seven producers to create new charts. Various comparisons (intended vs. perceived messages, initial vs. created charts, and within and between comparisons of the workshops and interviews) are conducted throughout the study.}
\end{figure*}

\section{Methodology}
\label{sec:methodology}
We conducted a series of workshops through extended focus groups and interviews to investigate how visualization producers encode messages about data and how visualization consumers engage with and understand these data visualizations (see \autoref{fig:method_process}). We collaborated with professional chart producers from diverse institutions who provided us with charts that they created and published. A sample of eight charts was shown to diverse consumer groups, and they were asked to write down what they thought each chart's main message was, individually and collaboratively. For collaborative sensemaking, we split participants into small groups, let them discuss the chart, and write down a single, collaboratively composed message. Additionally, we let them rate the charts based on different subjective dimensions (explainability, understandability, aesthetics, design, readability, clarity of terms, trustworthiness, and appropriateness; see \autoref{tab:likert_questions}) to allow for comparisons between the charts. We conducted five workshops and ten interviews with different participant groups, some of which are rarely considered in the academic visualization community, including senior citizens, university students, job seekers, designers, and young people from underprivileged backgrounds. We report on the qualitative and quantitative analysis in~\autoref{sec:results}.

Informed by the findings from the consumer workshops and interviews, we conducted a follow-up workshop with our collaborating chart producers. The workshop contained a series of activities where the participants were asked to sketch charts based on messages consolidated from the collected consumer messages. The goal was to compare the chart from our sample (initially provided to us by the chart producers) to the chart created during the producer workshop to see whether elements (e.g., type of chart, level of detail) changed and, if so, to see what caused these changes.

\subsection{Structure}
\label{sec:structure}

\begin{table}
    \centering
    \caption{Overview of our five workshops and ten interviews with 42 chart consumers.}
    \bgroup
    \renewcommand{\arraystretch}{1.2}
    \resizebox{0.4\columnwidth}{!}{\begin{tabular}{ clcl }
      \hline
      \textbf{ID} & \textbf{Consumer Group} & \textbf{\#Participants} & \textbf{Length}\\
      \hline
      W$_{1}$ & [U] University students & 4 & 3 hours \\
      W$_{2}$ & [U] University students & 9 & 1.5 hours \\
      W$_{3}$ & [S] (School) students & 13 & 3 hours \\
      W$_{4}$ & [J] Job seekers & 3 & 3 hours \\
      W$_{5}$ & [D] Designers & 3 & 2 hours \\
      \hline
      I$_{1-10}$ & [R] Retired people & 10 & 0.5 hours \\
      \hline
    \end{tabular}}
    \egroup
    \label{tab:consumerworkshops}
\end{table}

\begin{figure*}[!t]
  \centering
  \includegraphics[width=\textwidth]{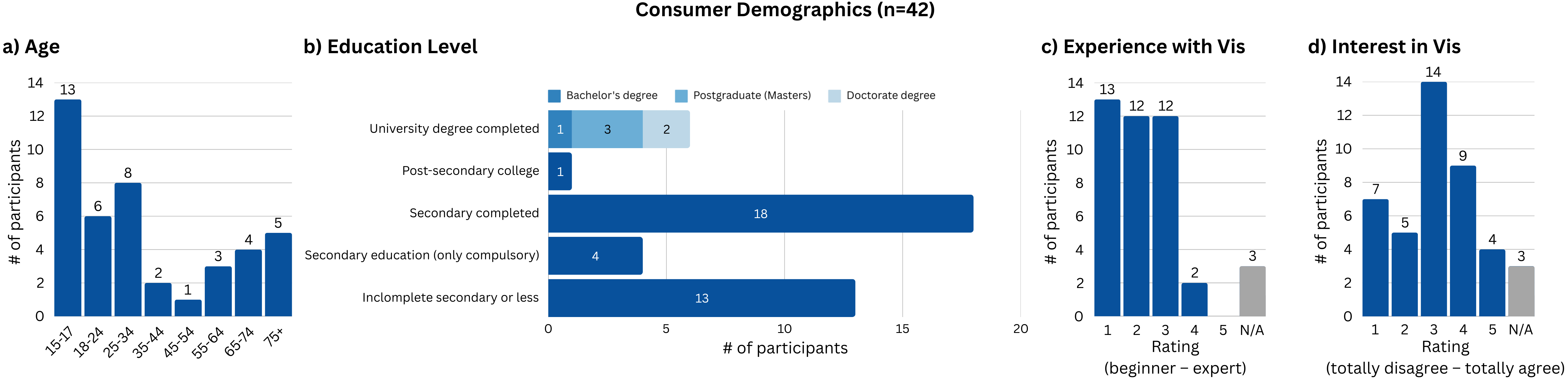}
  \caption{Demographic overview of our 42 participants in the consumer group. The bar charts show the distribution of the participants for the a) age range, b) education level, c) experience with visualization, and d) interest in visualization.}
  \label{fig:consumer_demo}
  \Description[Figure 2 shows the demographics of the 42 chart consumers.]{Figure 2 shows the demographics of the 42 chart consumers. There are four bar charts showing their age distribution, education level, experience with visualization (five-point Likert scale), and interest in visualization (five-point Likert scale).}
\end{figure*}

\textbf{Consumer workshop.}
We conducted five workshops over two months with different consumer groups (see \autoref{tab:consumerworkshops}). Each workshop took between 90 and 180 minutes. Through multiple consecutive activities, we explored what messages consumers take away from visualizations and how they rate the charts on different subjective dimensions. Each workshop started with an introduction to the project, an overview of the workshop structure, and a definition of data visualization since most participants had no prior visualization experience. In the first key activity, called ``What is the message?'', participants were shown one chart at a time and asked to note down the main message conveyed by the chart individually. They were not given a definition of a message since one of our objectives was to analyze how the messages differ in content and level of detail. Additionally, they filled out eight 5-point Likert scale questions to assess the visualizations subjectively (see \autoref{tab:likert_questions}). Afterward, the participants discussed their messages in small groups and collaboratively decided on a consolidated message.

In the final activity, participants had to write down what they liked and disliked about the charts, what made them trustworthy, and what they would change about them. They had 60 seconds for each question to ensure short, concise, and immediate answers. 

We conducted a pilot consumer workshop with four participants to test activities regarding their effectiveness and length. Based on the results, we adjusted minor parts of the workshop for more efficiency. Our final workshops ($W_{1}$--$W_{5}$) varied in length due to different factors (e.g., number of visualizations shown, time constraints), but we ensured that the time participants spent with a chart was equal for every workshop and interview so that the results were comparable.

\textbf{Consumer interviews.}
We opted to conduct interviews to gain the perspective of senior citizens, as, despite our best efforts, we were unable to recruit senior citizens for a workshop setting. We approached numerous retirement homes and social clubs, but arranging workshops was impossible because of their institutional regulations. Hence, we conducted ten individual interviews with senior citizens. The content of these interviews was as similar as possible to the workshops: articulating the chart's message, filling out the Likert questionnaire, and completing the subjective assessment for the chart. However, we only showed four of the eight visualizations per session to restrict the length of the interviews and did not create collaborative messages for this group. We did, however, discuss the charts with them, focusing on the same topics as in the workshop: like/dislike, trustworthiness, and changes to the chart.

\begin{table}
    \centering
    \caption{Chart producers' demographics.}
    \bgroup
    \renewcommand{\arraystretch}{1.2}
    \resizebox{0.6\columnwidth}{!}{\begin{tabular}{ clllll }
      \hline
      \textbf{ID} & \textbf{Gender} & \textbf{Age} & \textbf{Job Description} & \textbf{Work Experience} & \textbf{Education Level}\\
      \hline
      P$_{1}$ & female & 42 & Communication officer & 10 years & Postgraduate (Masters) \\
      P$_{2}$ & female & 40 & Graphic designer & 10 years & Postgraduate (Masters) \\
      P$_{3}$ & male & 43 & Infographic designer & 14 years & Post-secondary college \\
      P$_{4}$ & male & 54 & Head of infographics & 33 years & Postgraduate (Masters) \\
      P$_{5}$ & male & 36 & Infographic designer & 12 years & Secondary completed \\
      P$_{6}$ & female & 34 & Infographic designer & 5 years & Postgraduate (Masters) \\
      P$_{7}$ & male & 42 & Economist & 6 years & Doctorate degree \\
      \hline
    \end{tabular}}
    \egroup
    
    \label{tab:producers}
\end{table}

\subsection{Participants}
\textbf{Chart producers.}
We recruited seven professional chart producers (3 female, 4 male) between 34 and 54 years of age from diverse institutions through purposive sampling to ensure diversity in terms of gender and work experience (see \autoref{tab:producers}). They produce charts for the general public and have different job roles with a mean work experience of around 13 years ($\sigma$ = 9 years). We conducted a voucher raffle at the end of the workshop as an incentive for their participation. Additionally, we provided them with a report on the consumer feedback on their charts. All producers gave written consent to participate in the workshop.

\textbf{Chart consumers.} One of our main goals in recruiting a diverse participant sample was to include populations with varying socioeconomic profiles that are sometimes underrepresented in visualization research~\cite{lee2020,jena2021,while2024} or behavioral sciences~\cite{heinrich2010}. We targeted different consumer groups for our workshops and interviews: young students from underprivileged backgrounds (S), university students (U), retired senior citizens (R), job seekers (J), and designers (D). We recruited a heterogeneous sample of 42 participants between 15 and 95 years of age, aiming for a balanced gender ratio (self-reported gender: 18 female, 21 male, 3 not specified). Participants completed a demographic questionnaire at the end of each session. \autoref{fig:consumer_demo} shows an overview of consumer demographics. While most of our participants completed or were still in secondary education, we only had 6 participants with a university degree, which matched our goal of recruiting participants from a broader population segment. We specifically targeted consumers without prior visualization experience. Our final group of participants had little or no visualization experience except for the designer group. 
All participants received minor incentives such as catering during the workshop or interview, plus a small thank-you in the form of a gift, a voucher raffle, or minor course credits where applicable. The types of incentives differed mainly due to differences in regulations of the different institutions but were selected to be of similar value. The studies were carried out as part of a project that has undergone ethical screening according to the guidelines of our institution and was determined to be low-risk. All participants agreed to be (audio) recorded and gave written consent to participate in the study before each session. These recordings were later transcribed and analyzed. Additionally, a demographic questionnaire was completed at the end of each session.

In the following sections, we reference our consumer participants by a unique ID consisting of the first letter of the group name followed by a number (e.g., R-4 refers to participant four in the group of retired people).

\subsection{Chart Sample}

\begin{figure*}[!t]
  \includegraphics[width=\textwidth]{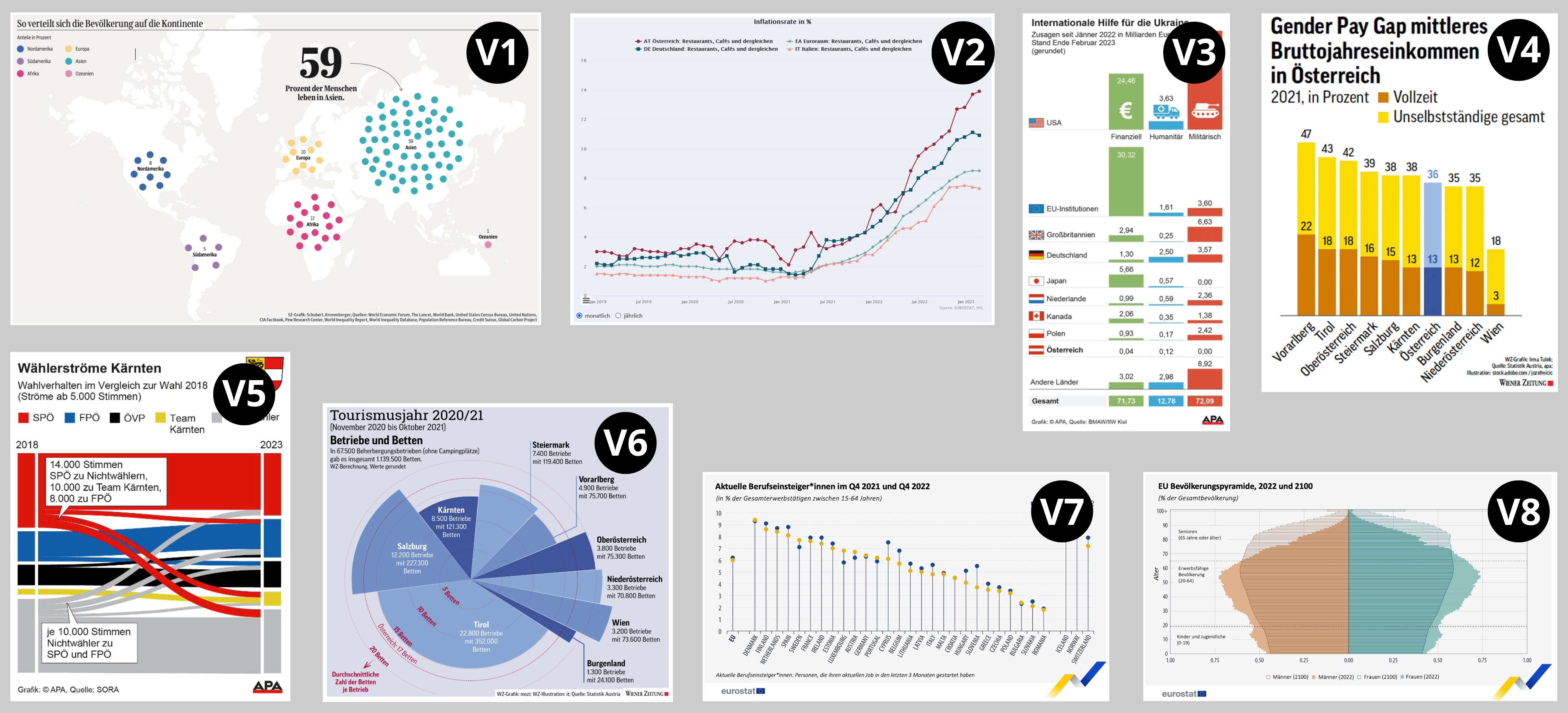}
  \caption{Sample of charts used in the workshops and interviews (see supplemental material S02 for full-page figures). \textit{Reproduced with permission. V$_{1}$: Graphic: Süddeutsche Zeitung (SZ from November 4, 2022) / Julia Schubert and Isabel Kronenberger; V$_{2}$: Graphic: IHS Preismonitor (2022) by Sebastian P. Koch and Christine Lietz, Source: Eurostat, IHS; V$_{3}$: Graphic: ©APA / Walter Longauer, Source: BMAW / IfW Kiel; V$_{4}$: Graphic: Wiener Zeitung / Irma Tulek, Source: Statistik Austria, APA; V$_{5}$: Graphic: ©APA / Walter Longauer, Source: SORA; V$_{6}$: Graphic: Wiener Zeitung / Moritz Ziegler, Source: Statistik Austria; V$_{7}$, V$_{8}$: Graphic, Source: Eurostat.}}
  \label{fig:charts}
  \Description[Figure 3 shows the chart sample and displays the eight charts of the sample.]{Figure 3 shows the chart sample and displays the eight charts of the sample. Each chart has a unique identifier.}
\end{figure*}

\begin{table*}
    \centering
    \caption{Overview of our chart sample with the visualization types according to Chen et al.~\cite{chen2022}, chart type and description of the chart content. V$_{1}$ through V$_{4}$ were labeled as \textit{simple} charts, whereas V$_{5}$ through V$_{8}$ were labeled as \textit{complex} charts by the chart producers.}
    \bgroup
    \renewcommand{\arraystretch}{1.2}
    \resizebox{\textwidth}{!}{\begin{tabular}{ clll }
      \hline
      \textbf{Vis-ID} & \textbf{Visualization Type} & \textbf{Chart} & \textbf{Description of the Content} \\
      \hline
      V$_{1}$ & Point-based Representation & Dot plot & Population distribution across the continents. \\
      V$_{2}$ & Line-based Representation & Line chart & Inflation rate in three selected countries and the euro area.  \\
      V$_{3}$ & Generalized Bar Representation & Grouped bar chart & Commitment for international help from selected countries for Ukraine. \\
      V$_{4}$ & Generalized Bar Representation & Stacked bar chart & Gender pay gap regarding the gross annual income in a selected country. \\
      \hline
      V$_{5}$ & Node-link Trees/Graphs, Networks, Meshes & Sankey diagram & Voter flow analysis of elections in 2018 and 2023 in a selected country. \\
      V$_{6}$ & Generalized Area Representation & Pie chart & Distribution of accommodations and beds (tourism data) in a selected country.  \\
      V$_{7}$ & Point-based Representation & Dot plot & Recent job starters in selected countries in subsequent years (2021 and 2022). \\
      V$_{8}$ & Generalized Bar Representation & Bar chart/Histogram & EU population pyramid in 2022 and 2100 (prediction). \\
      \hline
    \end{tabular}}
    \egroup
    \label{tab:charts}
\end{table*}

\begin{figure*}
  \centering
  \includegraphics[width=\textwidth]{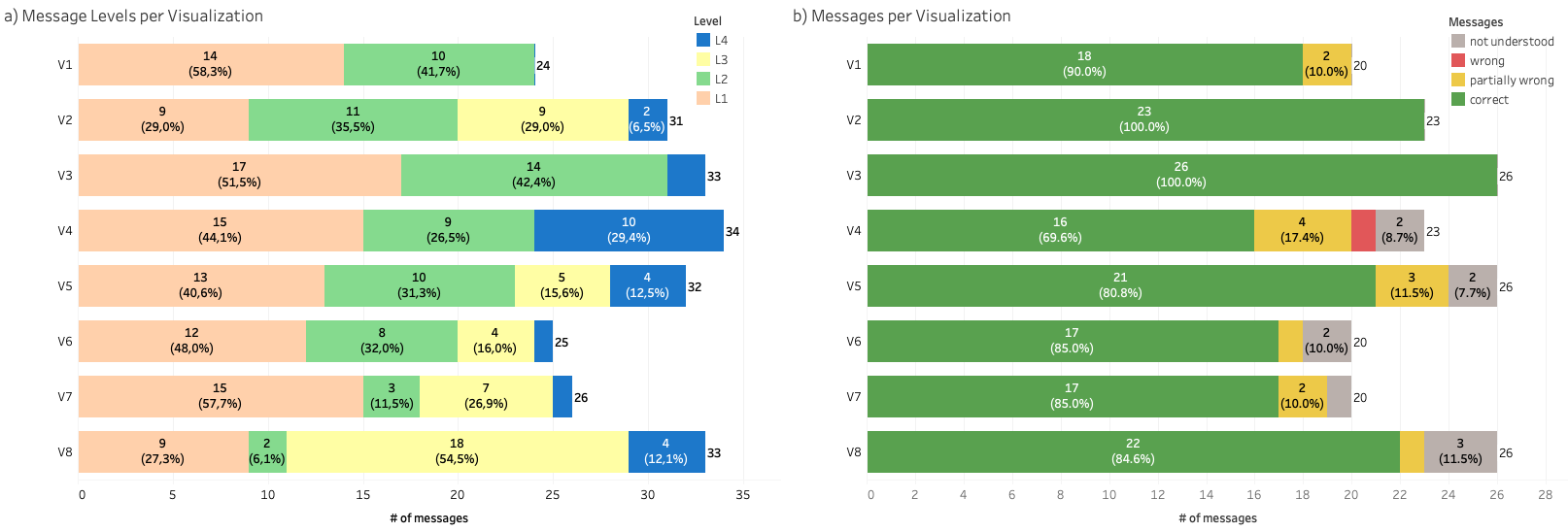}
  \caption{Typology of the 174 individual messages created by the consumer groups. Messages are grouped into the four different levels of messages (a) defined by Lundgard and Satyanarayan~\cite{lundgard2022} and based on the correctness of their content (b).}
  \label{fig:messages_bars}
  \Description[Figure 4 shows two bar charts comparing message levels and correctness across eight visualizations (V1 to V8).]{Figure 4 shows two bar charts comparing message levels and correctness across eight visualizations (V1 to V8). Chart A displays the distribution of messages across four levels (L1 to L4) as defined by Lundgard and Satyanarayan. Each bar (representing one sample chart) shows each level’s count and percentage. Chart B shows the correctness of messages, categorized as correct, partially wrong, wrong, or not understood, with green indicating correct responses and other colors representing varying degrees of error.}
\end{figure*}

We collaborated with seven professional chart producers who create charts for the general public on popular topics, including but not limited to demographics, elections, and recent political events. We asked them to provide us with at least one of their recent charts that is \textit{simple} and one that is a bit more \textit{complex} to grasp based on their intuition and label them accordingly. We specifically focused on static charts primarily intended for print media (e.g., newspapers). From the 29 visualizations that we obtained, we selected a sample of eight visualizations (see \autoref{fig:charts} and \autoref{tab:charts}) that have different characteristics, four for each category (\textit{simple} vs. \textit{complex}). The goal was a mix of visualizations with different characteristics (e.g., chart type, level of detail) intended for the general public, predominantly in the news context, manageable within the scope of the workshops and interviews. All of the charts in this sample were published in print or online media and targeted toward the general public, which was one of our main selection criteria.

\subsection{Analysis}
All interviews and workshop discussions were recorded, transcribed, and translated afterward. Since most questionnaires were filled out on paper, we manually digitized them into a spreadsheet, which was the basis for our analysis. We used Python and Tableau to explore the quantitative data. The messages collected from the participants were analyzed thematically based on two different typologies \cite{koesten2023,lundgard2022}, which we explain in \autoref{subsec:messages}. We conducted a content analysis~\cite{hsieh2005} of the participants' free-text responses.

\section{Findings}
\label{sec:results}
First, we describe the results of our analysis of the consumer workshops and interviews with 42 participants. On a sample of eight visualizations, we investigated how consumers interpret and rate charts on different subjective dimensions. We present an analysis of the 174 collected individual and 39 consolidated messages, followed by the questionnaire results.
We then presented our findings based on the producer workshops with seven participants, where we observed how professionals create charts individually and in a collaborative setting. We conclude by comparing their re-designed charts to the original charts and present insights into the chart producers' professional experience.

\begin{figure}
  \centering
  \includegraphics[width=0.4\columnwidth]{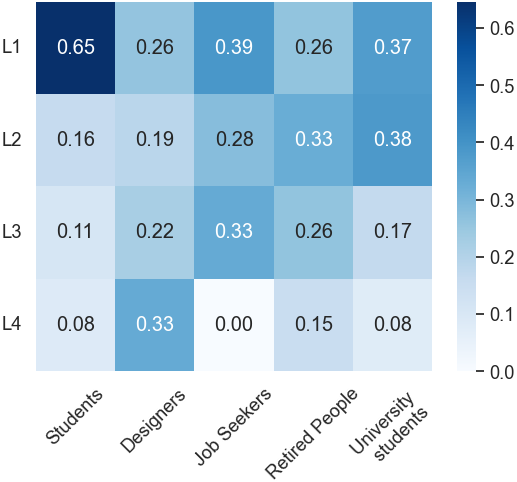}
  \caption{Heatmap showing the distribution of specific message types (L1-L4) within the different consumer groups.}
  \label{fig:heatmap_levels}
  \Description[Figure 5 is a heatmap showing the distribution of the four levels (L1 to L4) as defined by Lundgard and Satyanarayan across the five consumer groups.]{Figure 5 is a heatmap showing the distribution of the four levels (L1 to L4) as defined by Lundgard and Satyanarayan across the five consumer groups: Students, Designers, Job Seekers, Retired People, and University Students. A blue color scale on the right shows the value range from 0 to 0.6, and darker blue shades indicate higher values.}
\end{figure}

\subsection{Messages}
\label{subsec:messages}
We collected 184 individual messages and 39 consolidated messages for eight consumer charts. Ten of the individual messages were left blank; these messages indicated the inability of participants to formulate messages due to the difficulty of understanding these charts (see V$_{4-8}$ in \autoref{fig:messages_bars}a).
The result was a final sample of 174 individual messages with content. This section examines the types of messages we collected based on different message typologies.

\textbf{Typology.}
We used two typologies to analyze the content of the collected messages. The first typology, introduced by Lundgard and Satyanarayan~\cite{lundgard2022}, consists of a four-level model of the semantic content of visualizations. Level 1 (L1) refers to elemental and encoded properties of a visualization (e.g., visual components); level 2 (L2) refers to statistical concepts and relations (e.g., extrema, correlations); level 3 (L3) refers to perceptual and cognitive phenomena (e.g., complex trends, patterns); and level 4 (L4) refers to contextual and domain-specific insights (e.g., social and political context, explanations). Messages might not be exclusive to a single level but can consist of multiple levels. \autoref{fig:messages_bars}a) shows the distribution of levels for the messages of each visualization.
Except for V$_{1}$, V$_{3}$, and V$_{4}$, consumer messages for the other visualizations covered all four semantic content levels. While V$_{1}$ only covered L1 (58\%) and L2 (42\%), V$_{3}$ and V$_{4}$ were just missing L3.

We observed that participants usually describe the content of a chart in their message by referencing visual elements (e.g., the axis, data points) or using the text on the chart, as seen in the following message: ``\textit{The main message is about the population on the continents. Several continents are displayed. Each continent has a different color and a different percentage.}'' (S-2 on V$_{1}$). The share of L1 also reflects this: except for V$_{2}$ and V$_{8}$, L1 had the biggest share in the messages for all visualizations. This underlines findings by Koesten et al.~\cite{koesten2023}, who looked at charts with and without context, showing the importance of text on a chart in determining takeaway messages.

Interestingly, most of the L2 messages (statistical concepts) referred to extrema (minima and maxima) in the chart, for example: ``\textit{The highest military support for Ukraine comes from the US while most financial support comes from the EU.}'' (J-2 on V$_{3}$) or ``\textit{[...] The gender pay gap is lowest in Vienna and highest in Vorarlberg [...]}'' (U-4 on V$_{4}$). This is especially true for V$_{1}$--V$_{6}$, while V$_{7}$ and V$_{8}$ yielded more L3 messages that contained trends over time.

Our analysis shows that charts containing time as a dimension (V$_{2,7,8}$) had a large share of L3 messages. Most of these L3 messages focused on trends over time, e.g.,: ``\textit{In 2021, there were, on average, more job starters than in 2022.}'' (U-13 on V$_{7}$) or ``\textit{The graphic shows a prognosis that by 2100, there will be way fewer men \& women at working age, and there will be a massive increase of elderly people.}'' (D-2 on V$_{8}$).

L4 messages, which include participants' contextual knowledge and domain-specific expertise, were specifically common with V$_{4,5,8}$. The chart on the gender pay gap (V$_{4}$) did not explain what the term means. School students especially had trouble interpreting this chart. However, most participants could recall their own knowledge on this topic and assumed that it shows the difference between the income of men and women: ``\textit{Men, who are not self-employed, earn on average 36\% more than women, while male full-time employees earn 13\% more.}'' (U-12 on V$_{4}$). Some participants even added their geographical knowledge to the message to explain the pay gap: ``\textit{Austria's western regions are more conservative than the eastern ones.}'' (D-1 on V$_{4}$).
We also observed that L4 messages sometimes included personal knowledge on missing data (``\textit{[...] Isn't the green party missing in the chart?}'', D-1 on V$_{5}$) or future predictions (``\textit{The graphic shows [...] an increase of elderly people who we will have to take care of. This will be a problem for our current pension system.}'', D-2 on V$_{8}$).
 
Looking at the distribution of message levels per consumer group (see \autoref{fig:heatmap_levels}), we can see that 65\% of the students' messages were L1 messages, more than in any other group, and only a third of their messages related to the other levels. We observed that the students' messages often reproduced textual information on the chart (e.g., the chart's title) when they had difficulty 
interpreting it: ``\textit{EU population pyramid 2022 and 2100 (in \% of total population)}'' (S-2 on V$_{8}$). Interestingly, one-third of the messages in the designer group related to L4. As shown above, messages were often complemented with personal knowledge about the chart's topic. On the contrary, job seekers did not write L4 messages; they were focused on extrema and trends in the data and the visual elements in the chart. This was very similar to the group of retired people, but 15\% of their messages also contained L4: ``\textit{The price increase in Austria is above the average for different reasons (e.g., COVID-19, war in Ukraine)}'' (R-5 on V$_{2}$). University students mostly wrote messages classified as L1 and L2: ``\textit{Comparison of the inflation rate between AT \& neighboring countries. AT has the highest inflation rate, while IT has the lowest inflation rate. [...]}'' (S-1 on V$_{2}$).

To summarize, the distribution of the different message types varied across the different consumer groups, mainly focusing on visualization properties (L1), statistical information (L2), and perceptual and cognitive phenomena (L3). Interestingly, we observed that students who did not have much prior knowledge of the topics presented in the charts focused mainly on visualization properties (L1) in their messages, indicating that this missing topic knowledge might lead to messages that recall parts of the charts. On the other hand, designers who work daily on a wide range of topics were much more open and invested in also including textual and domain-specific insights (L4).

We further analyzed the messages according to Koesten et al.'s~\cite{koesten2023} typology, describing messages by their granularity (general, mid-level, specific; exclusive), their composition (if they contain examples, descriptions, or nonsense; not exclusive), and their structure (additive, simple; exclusive). Almost 43\% of all individual messages were very general without any details about the data, 39\% contained some level of detail, and 18\% included longer, very detailed text. General messages usually consisted of a single semantic level (L1--L4), while the others often contained multiple levels. 59\% of the collected individual messages contained descriptions of visual elements closely related to L1, and half of all messages (50\%) also contained specific examples for data points. Only around 1\% of the messages were classified as ``nonsense'' as participants only stated the overall topic of the chart without any further explanation: ``\textit{It [remark: the chart] is about an election}'' (S-11 on V$_{5}$). Two-thirds of the messages contained a single and simple message, while the other third consisted of multiple small messages adding to a summative message.

\textbf{Correctness.}
Checking the ``correctness" of the content of each message is an indicator of whether a specific chart or its corresponding chart type might be misleading (see \autoref{fig:messages_bars}b).
We determined correctness based on our reading of the charts as visualization researchers. While the first three charts (V$_{1}$--V$_{3}$) mainly produced correct messages, two retired participants (R-6, R-9) did not read the title of V$_{1}$ and interpreted the chart solely by its visual elements: ``\textit{Asia has the highest population growth}'' (R-6 on V$_{1}$) and ``\textit{Spread of a disease over the continents.}'' (R-9 on V$_{1}$).   
Looking at V$_{4}$, only around 70\% of the messages regarding the stacked bar chart about the gender pay gap were semantically correct, particularly the retired and student groups had difficulty interpreting this chart. This mainly was because participants did not know the meaning of the gender pay gap, resulting in (partially) wrong messages: ``\textit{Average gross annual income in Austria}'' (R-10 on V$_{4}$). An explanation of the term could have helped participants grasp the information shown in the chart. V$_{5}$ shows a Sankey diagram about elections in two different years, a prevalent chart type for this type of information. However, two participants (S-10, J-2) did not understand the chart, and three others (S-12, R-2, J-3) produced partially wrong messages. The crossing lines and the additional encoding of the number of voters through line thickness confused them. The visual overload was also mentioned as an issue in V$_{6}$--V$_{8}$, yielding only around 85\% correct messages. 

To summarize, we could observe that participants had particular difficulties interpreting charts that have a high number of data points (V$_{7}$, V$_{8}$) or dimensions (V$_{6}$, V$_{8}$), or have missing descriptions of domain-specific terms used on the chart (V$_{4}$). Additionally, interpreting unconventional graphical depictions (V$_{1}$, V$_{6}$,  V$_{7}$, V$_{8}$) was often considered challenging.

\textbf{Collaborative Messages.}
We collected 39 collaborative messages in our consumer workshops. Participants were asked to discuss their messages in groups and to work together to develop a consolidated message, which we term a ``collaborative message'' informed by collaborative sensemaking~\cite{mahyar2014}. We observed participants during this process. For most groups, consensus was reached after a discussion where everybody contributed a fair share. In some groups (e.g., the school students), a few participants were driving the conversation in each group. In contrast, others were focused on listening and agreeing and, therefore, did not actively contribute. Especially in the school student group, we observed that some participants were constrained by the German language as they had different German proficiencies since most of them had another native language. In most cases, the collaborative messages were either shorter or had the same length as the individual ones, focusing on a specific aspect of the chart's data and condensing the message content: ``\textit{There will be more elderly people than younger ones.}'' (U-5 through U-9 on V$_{8}$). For some of the charts (mostly the ones labeled by the chart producer as being easy), people were able to come up with the consolidated message quickly, while other charts (V$_{4}$, V$_{5}$, V$_{6}$, and V$_{7}$) took longer as the information density was relatively high. Especially V$_{4}$ (the stacked bar chart) was challenging for some participants as they had varying interpretations of the numbers on the chart. However, in all cases, participants eventually agreed on a consolidated message. Using their shared knowledge, some discussions led to new information not represented in the chart. When discussing the gender pay gap, the designers integrated their geographical knowledge: ``\textit{It appears that towards the western regions, the gender pay gap would increase in percentage, right? This is the main point I can see here. It means go west, young man. Or go east, young woman.}'' (D-1 through D-3 on V$_{4}$). However, despite our prior assumption that the collaborative sensemaking process would develop message content not covered in the individual messages, we did not observe this in our study. Most of the collaborative messages were condensed summaries of the individual messages, while others were also more high-level than the individual messages, focusing on the overall topic of the chart rather than details or interpretations.

\begin{table*}[h]
\centering
\caption{Examples for alignment of consumer messages with intended messages.}
    \renewcommand{\arraystretch}{1.2}
    \resizebox{\textwidth}{!}{\begin{tabular}{m{2.8cm}|m{14.2cm}}
    \hline
    \multicolumn{2}{c}{\textbf{Alignment of selected consumer messages with an intended message (examples: $V_{3}$ and $V_{6}$)}} \\ 
    \hline
    \rowcolor[HTML]{EFEFEF} 
    \multicolumn{2}{p{17.5cm}}{\textbf{Intended message for $V_{3}$:} Overview of the most important donor countries for aid to Ukraine compared to the EU, with a focus on military aid from the USA.} \\ 
    \hline
    \textbf{High alignment} & Shows aid to Ukraine, much of which comes from the USA (divided by country). (R-2) \\ 
    \hline
    \textbf{Medium alignment} & Distribution of aid to Ukraine since January 2022 from other countries in the financial, humanitarian, and military areas. (U-7) \\ 
    \hline
    \textbf{Low alignment} & Most of the money goes to military support. Humanitarian aid would be the most important, but that is the smallest. (R-3) \\ 
    \hline
    \rowcolor[HTML]{EFEFEF} 
    \multicolumn{2}{p{17.5cm}}{\textbf{Intended message for $V_{6}$:} The goal was to illustrate which federal state offers the highest number of beds in lodging establishments (represented by pie chart segments) and which state has, on average, the largest accommodations. In Vienna, it can be seen that there are few beds overall but the most beds per establishment.} \\ \hline
    \textbf{High alignment} & Tyrol has the most accommodations with 22,800 businesses. Burgenland has a lot of beds in relation to the number of businesses. Vienna has the most beds per accommodation. (U-2) \\ \hline
    \textbf{Medium alignment} & The graphic shows the shares of accommodation businesses and beds in the individual federal states compared to the whole of Austria. Tyrol has the most beds and establishments. (U-10) \\ \hline
    \textbf{Low alignment} & It's about accommodation businesses and beds in Austria. (S-6) \\ \hline
    \end{tabular}}
    \label{tab:intended_message}
\end{table*}

\textbf{Intended Message.}
In addition to their charts, the chart producers provided us with the intended message for each chart which can be found in the supplemental material (S01). Similarly to the consumers' messages, we did not prompt producers or provide any format or structure for the intended message as we wanted to see how they would interpret and convey the takeaway of a chart. While the intended messages for six of the charts (V$_{1}$--V$_{6}$) were rather condensed and specific (e.g., ``\textit{Overview of the most important donor countries for aid to Ukraine compared to the EU, with a focus on military aid from the USA.}''; V$_{3}$), the other two were longer and also contained the producers' reasons on why they visualized this data (e.g., ``\textit{Given the impact of COVID-19 on the labor market, one aspect was to show if the labor market situation is "recovering" by showing the development. [...]}''; V$_{7}$). We compared the intended messages to the consumers' messages, which we see as multidimensional constructs consisting of different facets (e.g., focus and level of abstraction). Thus, we compared them qualitatively and found that, in most cases, the essential, intended message was understood, but some details differed.
The varied nature of the messages, which can take different forms ~\cite{koesten2023} calls for the use of in-depth, qualitative analysis of alignment. We, therefore, illustrate the alignment in \autoref{tab:intended_message} using example messages.

All charts were created to give an overview and show the data distribution, e.g., for comparing different data points. We observed that comparisons were often used to describe extrema in the data: ``\textit{The gender pay gap in Vienna is far below the Austrian average. In Vorarlberg, it is almost 50\%.}'' (U-2 on V$_{4}$). More than half of the charts (V$_{1}$, V$_{3}$, V$_{4}$, V$_{5}$, V$_{6}$) aimed to communicate extrema, which was reflected by most of the participants' messages: ``\textit{Tyrol has the most accommodations \& beds in 2020/21, with an average of 15 beds per accommodation while Vienna has the largest average number of beds per accommodations. [...]}'' (U-1 on V$_{6}$). For one chart (V$_{7}$), the intended message was to show the development of a specific measure during the global COVID-19 pandemic. However, none of the participants' messages included this context since the chart did not mention the pandemic (e.g., through text on the chart).

\subsection{Subjective Measures}
\label{subsec:subjectivemeasures}

\begin{table}
    \centering
    \caption{5-point Likert scale questions (totally disagree -- totally agree) filled out by the consumers for the charts and their connection to different characteristics that target subjectively perceived skills.}
    \bgroup
    \renewcommand{\arraystretch}{1.2}
    \resizebox{0.7\columnwidth}{!}{\begin{tabular}{ cll }
      \hline
      \textbf{ID} & \textbf{Question} & \textbf{Connected to}\\
      \hline
      Q$_{1}$ & I could explain the chart's content to someone else. & Explainability \\
      Q$_{2}$ & The chart is understandable. The information can be read clearly. & Understandability \\
      Q$_{3}$ & The chart is nice to look at. & Aesthetics \\
      Q$_{4}$ & The chart is well designed. & Design \\
      Q$_{5}$ & The chart is readable. & Readability \\
      Q$_{6}$ & The chart only uses words I know. & Clarity of terms \\
      Q$_{7}$ & I think the chart is trustworthy. & Trustworthiness \\
      Q$_{8}$ & I would not have used a different visualization for the information shown. & Appropriateness \\
      \hline
    \end{tabular}}
    \egroup
    \label{tab:likert_questions}
\end{table}

\begin{figure*}
  \centering
  \includegraphics[width=\textwidth]{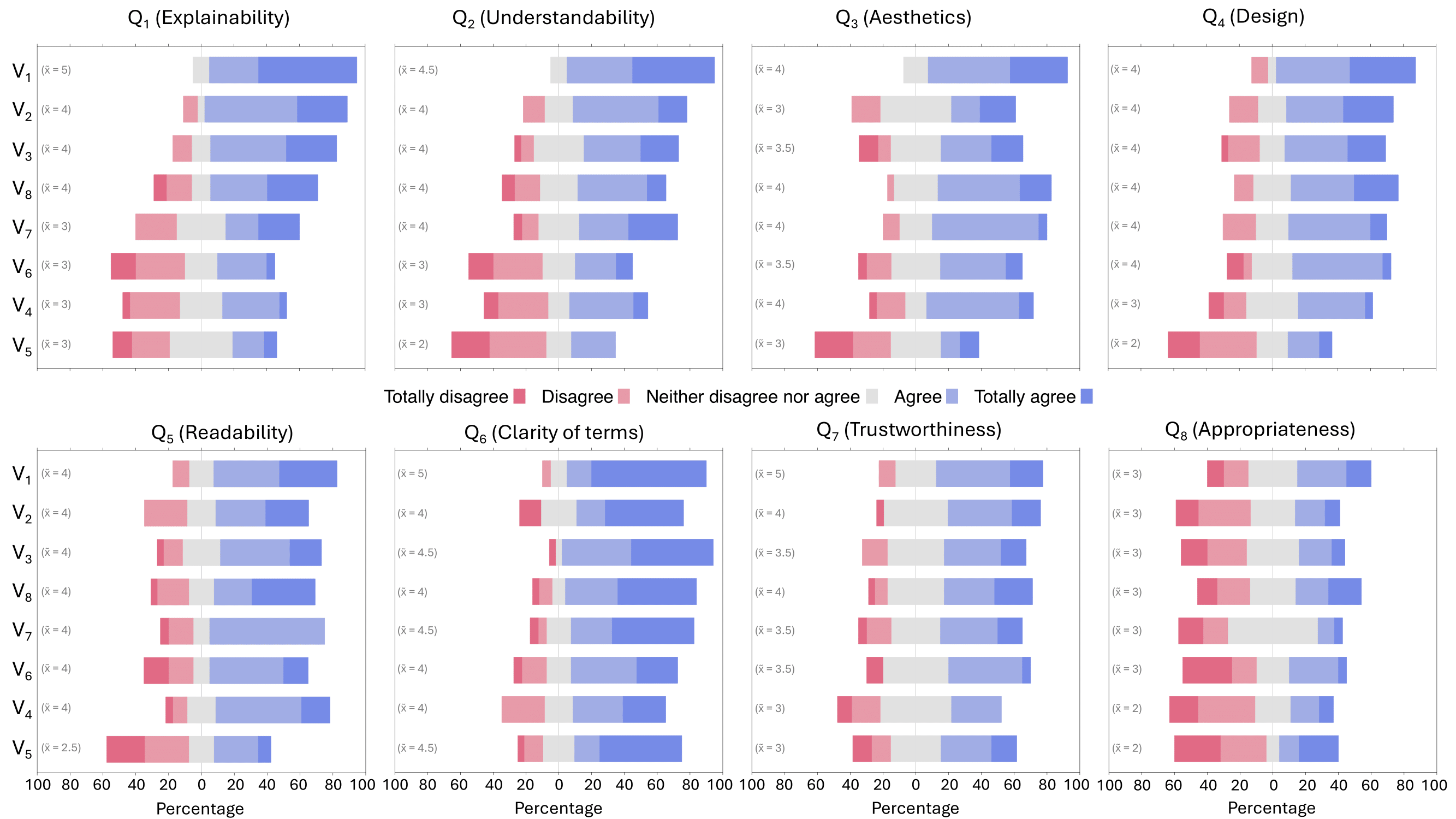}
  \caption{Overview of the results of the Likert questions for each visualization with the median rating ($\tilde{x}$; 1 corresponds to the rating \textit{totally disagree}, and 5 corresponds to \textit{totally agree}). 
  The visualizations are ordered from best-rated ($V_{1}$) to worst-rated ($V_{5}$) overall. \textit{Remark: not all visualizations were shown to all consumer groups.}}
  \label{fig:likert_results}
  \Description[Figure 6 shows the eight charts for the Likert questions, displaying the Likert scores for each visualization as a divergent stacked bar chart with five colored bar segments (totally disagree, disagree, neither disagree nor agree, agree, totally agree).]{Figure 6 shows the eight charts for the Likert questions, displaying the Likert scores for each visualization as a divergent stacked bar chart with five colored bar segments (totally disagree, disagree, neither disagree nor agree, agree, totally agree). Each chart corresponds to a question (Q1–Q8) on topics like explainability, understandability, aesthetics, design, readability, clarity of terms, trustworthiness, and appropriateness. The median for each visualization and question is shown as text next to the visualization name.}
\end{figure*}

Participants were given a questionnaire containing eight 5-point Likert questions about different subjective dimensions (see \autoref{tab:likert_questions}) for each chart. 
We compared the charts by the participants' ratings for each measure. Additionally, participants were asked four open-ended questions about each chart, for which they had 60 seconds to answer to ensure concise and immediate answers. These complemented the Likert ratings, including what they liked and disliked about the chart, what makes it trustworthy, and what they would change about it.
We conducted a content analysis~\cite{hsieh2005} of the 564 answers to these questions provided by 42 participants.
Depending on the workshop or interview structure, not all participants answered the questions for all charts.

\textbf{Overall ratings.}
In the following section, the percentage in the parentheses indicates the share of all participants that mentioned the corresponding theme. 
\autoref{fig:likert_results} shows a quantitative summary of the consumers' ratings for each Likert question and visualization, ordered from the overall best-rated visualization (V$_{1}$) to the worst-rated visualization (V$_{5}$). V$_{1}$ (dot plot showing the population distribution across the continents) immediately appears to have the best Likert scores from all visualizations for all categories. Participants specifically liked the colors (48\%) and simplicity (29\%) of the chart and the use of points as visual marks on a map (19\%). However, improving some visual elements (38\%) was also suggested, including bigger font sizes, using an isotype representation, and thicker continent contours. 
V$_{4}$, V$_{5}$, and V$_{6}$ have mostly lower ratings than the other visualizations. V$_{4}$ got low ratings for all measures except aesthetics (Q$_{3}$) and readability (Q$_{5}$). Half of the participants found the chart confusing (53\%) due to the missing explanation of the term `gender pay gap.' Additionally, 37\% did not like the color and would use a different chart type. This aligns with our previous analysis of the V$_{4}$ messages. In most cases, the designer group gave lower scores than the others (see supplement S01).
While V$_{5}$ got low ratings for clarity of terms on the chart (Q$_{6}$) and appropriateness (Q$_{8}$), it scored even lower in all other dimensions. Students and university students, in particular, gave lower scores on this chart. 80\% of the participants were confused by the Sankey diagram's flows and crossings. While Sankey diagrams might seem useful for electoral data as they represent static charts of dynamic processes~\cite{riehmann2005}, the results suggest they are challenging to understand and might lead to confusion.
To alleviate this problem, participants suggested adding more textual information on the chart (53\%) or using a different chart type (50\%). However, 70\% of the participants liked the colors in the Sankey diagram as they represent the official colors of the individual political parties.
The participants' main issue with V$_{6}$ was the confusing data representation (56\%) and the overload of information (28\%). The chart used multiple dimensions for encoding the data, which was perceived as overwhelming. Interestingly, despite that, 22\% of the participants liked the chart's aesthetics, which suggests that the effectiveness of a chart might not affect aesthetic perception.

To summarize, we found that participants had difficulties interpreting charts that have a high information density (V$_{5}$, V$_{6}$, V$_{7}$) or show multiple dimensions (V$_{6}$). Another problem arose from unknown terms on the chart without explanation (V$_{4}$). The most liked visual element on the charts was color (28\%). 
In most cases, an indication of the data source (40\%) or a validation of the data through general knowledge (11\%) resulted in high trustworthiness ratings. The most suggested modification for a chart was selecting a different chart type (22\%).

\textbf{Comparing \textit{easy} and \textit{complex} charts.}
We performed Mann-Whitney U tests ($\alpha$ = 0.05) on our ordinal Likert data to compare the ratings for the \textit{easy} (V$_{1}$--V$_{4}$) and \textit{complex} (V$_{5}$--V$_{8}$) labeled charts. We combined the ratings of all visualizations in each group for each Likert question and compared the two groups. The results did not show significant differences between the ratings for Q$_{3}$ (Aesthetics; U = 4731.0, p = 0.15), $Q_{6}$ (Clarity of terms; U = 4388.0, p = 0.55), $Q_{7}$ (Trustworthiness; U = 4255.0, p = 0.95), and $Q_{8}$ (Appropriateness; U = 4083.0, p = 0.92). However, the results show significant differences between the \textit{easy} and \textit{complex} charts for Q$_{1}$ (Explainability; U = 5560.0, p = 0.00014), Q$_{2}$ (Understandability; U = 5440.5, p = 0.00053), Q$_{4}$ (Design; U = 4895.5, p = 0.039), and Q$_{5}$ (Readability; U = 4967.0, p = 0.034), indicating better ratings for the \textit{easy} charts in those aspects. These results align with our qualitative findings outlined above.

\subsection{Producer Workshop}
\label{subsec:producerworkshop}

Using the findings from the consumer workshop, we conducted a follow-up workshop with the chart producers. We split the workshop into two online sessions on different days, totaling about five hours.

First, participants individually sketched a chart to convey a given message (see \autoref{tab:producer_messages}) drawn from the collected consumer messages (referring to two \textit{easy} and two \textit{complex} charts). These charts were presented and discussed in the group. Participants were then provided with the data behind the selected message (in a table) and discussed whether they would change something about their chart. This process was repeated for three more messages.

Participants were further asked to finish four sentences regarding data visualizations to learn more about their practice (see last paragraph \textit{Professional experience} at the end of \autoref{subsec:producerworkshop}): ``A data visualization is aesthetic/well designed/trustworthy when...'' and ``A data visualization can help with...''.

Participants collaboratively designed a consolidated version of the chart in the second session based on their individual sketches. Since we had to restrict the length of the workshop, collaborative charts were designed for M$_{1}$ and M$_{4}$ as these showed the greatest variety of individually created charts. All chart sketches can be found in the supplemental material. Next, we report on the chart iteration process to illustrate how participants collaboratively engaged with, made sense of, and converted the message into a chart.

\begin{table*}[t]
    \centering
    \caption{The four messages provided to the chart producers in the producer workshop. Based solely on these messages, chart producers created charts individually as well as collaboratively.}
    \renewcommand{\arraystretch}{1.2}
    \resizebox{\textwidth}{!}{\begin{tabular}{cm{15cm}c}
      \hline
      \textbf{ID} & \textbf{Textual Description} & \textbf{Based on}\\
      \hline
      M$_{1}$ & The inflation rate in Austria, Germany, and Italy rose sharply from January 2019 to January 2023. Except for Italy, every country is above the euro area's average inflation rate. & V$_{2}$ \\
      \hline
      M$_{2}$ & Ukraine receives aid money (divided into the areas: financial, humanitarian, and military) with the US \& EU institutions supporting Ukraine the most in the area of financial and military aid money from the US being the highest. & V$_{3}$ \\
      \hline
      M$_{3}$ & A comparison of the results of a political party election in 2018 and 2023 shows that the share of votes for each party and for non-voters has remained similar. However, the voting behavior of voters has changed. & V$_{5}$ \\
      \hline
      M$_{4}$ & The population distribution by age in 2100 is changing compared to 2022, according to a forecast: the share (for both men and women) of children and youth (0-19) and the working-age population (20-64) is decreasing, while the share of senior citizens (65 and older) is increasing sharply. & V$_{8}$ \\
      \hline
    \end{tabular}}
    \label{tab:producer_messages}
\end{table*}

\textbf{Message 1.} 
Most participants decided to use a line chart to encode M$_{1}$ and only two (P$_{2}$, P$_{4}$) used a grouped bar chart. For the bar charts, the grouping of bars differs: while one (P$_{2}$) shows a bar for each country in a group representing one year, the other one shows two bars (for 2019 and 2023, respectively) grouped by country. Participants argued that the message mentions two specific time points, leaving it for interpretation of how many steps in between should be shown. Others wanted to show the evolution of inflation over time rather than using discrete time points, which was their main argument for using a line chart. While most used a single line chart containing multiple lines (one for each country and the euro area) to show the development, one participant (P$_{1}$) decided to use small multiples, one for each country: ``\textit{Based on my experience, users often want to only look at a specific country (e.g., their own) which is why I decided to use three line charts instead of a single one. It is also cognitively easy to compare these three countries across the different charts.}'' (P$_{1}$). 
After receiving example data (e.g., monthly inflation rates), most participants would have made minor adjustments to their charts, such as changing data granularity. They argued that M$_{1}$ could be split into two messages (inflation rate evolution vs. comparison to the euro area average), each requiring a different chart design.
\textbf{Message 2.}
The created charts for M$_{2}$ showed a greater chart type variety than for any other message. P$_{1}$ and P$_{6}$ used a pie chart divided into three areas (financial, humanitarian, and military), which were further split into countries and institutions. They mainly intended to show a part-to-whole visualization comparing the different areas. However, after sketching the chart, P$_{1}$ raised concerns: ``\textit{Now, looking at my chart again, I think it is too busy regarding accessibility. That is why I like discussing ideas with other chart designers to develop alternative designs.}'' To avoid visual overload in a single pie chart, P$_{2}$ opted for three pie charts, one for each area: ``\textit{I wanted to make the chart as readable as possible. This is why I decided to use three pie charts, which, as a whole, makes it complete.}'' The other participants used a stacked bar chart (P$_{3}$, P$_{7}$), a Sankey diagram (P$_{5}$), and a dot plot (P$_{4}$).
Participants had strong opinions on potential adjustments after seeing the data (which included more countries and total sums): ``\textit{My chart would have been significantly different. [...] By having information on more countries and also the total sum, I would have been tempted to create a pie chart or a treemap.}'' (P$_{4}$). P$_{1}$ and P$_{2}$ would change their pie chart to a stacked bar chart since it can be confusing if there are too many data points. P$_{1}$ added that  ``\textit{[...] the additional countries could be aggregated into a single category called `others', which would make the chart easier to grasp.}'', which was also encouraged by P$_{2}$ and P$_{5}$.

\textbf{Message 3.}
Six of the seven chart producers chose a Sankey diagram for M$_{3}$. P$_{2}$ emphasized its popularity for voter transition analysis -- a view shared by most participants. 
However, they also pointed out issues with this type of chart: ``\textit{Usually, I would avoid such static Sankey diagrams as they might overwhelm users. Since this is the most popular chart in the context of elections, I decided to use it. However, I would create an interactive version of this chart if possible.}'' (P$_{1}$). P$_{6}$ sketched three pie charts, each showing party shares with voter groups represented by differently colored layers.

\textbf{Message 4.}
Five participants used general bar representations (two grouped bar charts, one stacked bar chart, and two histograms) to encode the message visually. Except for P$_{7}$, all used the aggregated age groups from the message as actual data points, while P$_{7}$ used a population pyramid showing individual age shares. P$_{1}$ mentioned that ``\textit{[...] bar charts are a good and simple choice in this case as they allow for an easy comparison between the bars.}''. P$_{2}$ used a line chart to visualize the differences between the two years (2022 and 2100), and P$_{3}$ created a customized radial plot.
While the message only contained three age groups, the original data included the percentage for each age between 1 and 100+. This raised the question of how much granularity is needed to convey the same message and calls for zooming in/out functionalities. Based on this information, most participants would have used a more continuous data representation: ``\textit{Since there is data for every age, I would not have used the aggregated age groups but created a more continuous chart type (e.g., a population pyramid).}'' (P$_{6}$). P$_{1}$ said that based on the message, she would have assumed no explicit difference or distinction between women and men. P$_{5}$ additionally raised the question of how non-binary gender types could be represented.

\begin{figure*}
  \centering
  \includegraphics[width=\textwidth]{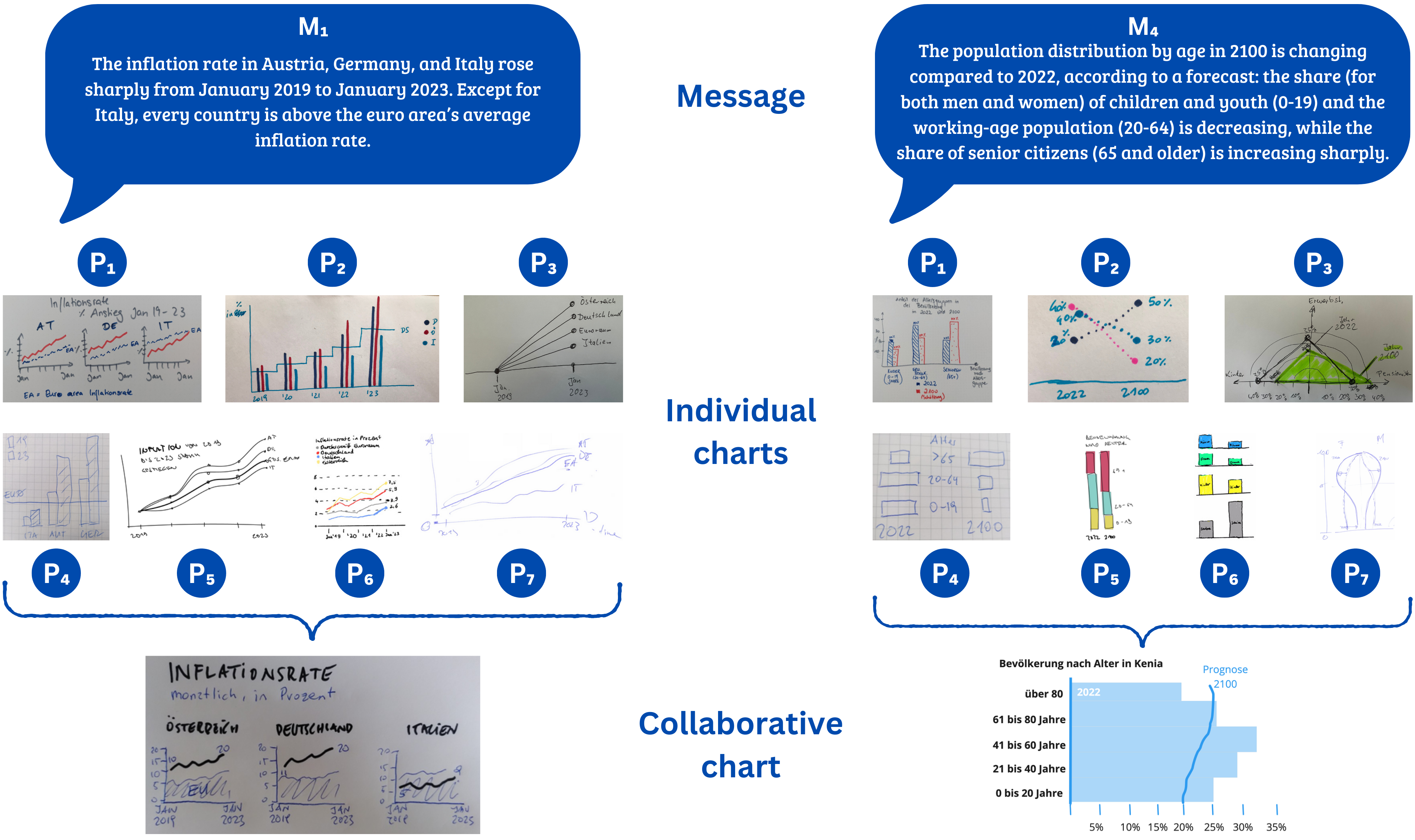}
  \caption{Participants created and discussed seven individual charts for each message in the first session of the producer workshop. This example shows the individual and corresponding collaborative charts for M$_{1}$ and M$_{4}$ from the second session. All the other charts can be found in the supplemental material S04.}
  \label{fig:created_charts}
  \Description[Figure 7 shows two messages, M1 and M4, from the producer workshop and the resulting seven individual charts and one collaborative chart for each message.]{Figure 7 shows two messages, M1 and M4, from the producer workshop and the resulting seven individual charts and one collaborative chart for each message. M1 discusses the sharp rise in inflation rates in Austria, Germany, and Italy from January 2019 to January 2023, with a collaborative chart showing monthly inflation rates for these countries. M4 addresses changes in the population distribution by age between 2022 and 2100, noting a decline in the working-age population and a rise in senior citizens with a collaborative chart showing age distribution forecasts for different age groups.}
\end{figure*}

\textbf{Collaborative charts.}
In the second session of the workshop, participants debated their respective charts and created a chart for M$_{1}$ and M$_{4}$ collaboratively (see \autoref{fig:created_charts}). Throughout this part, they discussed their professional experiences and production practices that inform their chart design.

For M$_{1}$, they discussed the pros and cons of bar charts versus line charts, the two main types used for their individual charts. Initially, they decided on a grouped bar chart showing countries as bars grouped by years, based on the chart of P$_{2}$. Regarding the orientation of the bars, P$_{4}$ argued: ``\textit{Vertical bar charts are intuitive if the x-axis is showing time or if we want to show increasing or decreasing trends. I usually use horizontal bar charts when the bar labels are too long.}'' After looking at the data, they decided to switch to a line chart instead: ``\textit{Having discrete bars for each data point could purport a continuous representation, and therefore, a line chart is the better option.}'' (P$_{3}$). P$_{2}$ added, it does not matter which of these two types to choose since both represent the message equally. The final design decision was a representation through three line charts, each representing a specific country, based on the chart of P$_{1}$. They adjusted this design by adding labels for the years on the x-axis: ``\textit{[...] it is not enough to put information, such as the time period, in the title; we also need to put it in the graph; otherwise, people will not understand it. [...] The title is an important teaser, but the chart needs to be understandable on its own.}'' (P$_{4}$). This aligned with a comment of P$_{2}$, who mentioned always looking at the chart before reading the title. The producers also talked about adding even more information to the chart. However, they decided against it so that the chart only conveys the information in the message, illustrating how a precise message focuses on design choices. They added the eurozone average inflation as an area in the background to allow comparisons.

For M$_{4}$, participants initially focused on a customized plot by P$_{3}$, which shows the data as a triangle (for each of the three age categories) in half of a radial plot (see \autoref{fig:created_charts}): ``\textit{I really like this plot because the triangles overlap making it easy to compare them.}'' (P$_{4}$). However, they agreed that this plot might be too complex for the general public.
They argued that these charts are often the starting point and support of a story, so when readers do not understand the chart, they might not continue reading an article. This was why they switched to a population pyramid, commonly used for showing population distribution. Even though the data contains information on all ages, they decided to aggregate them into age groups for simplicity. Additionally, since there is no significant difference in the distribution between women and men, they decided to use a single representation for both genders. They also decided against overlapping bars for the prediction of 2100: ``\textit{We represent the population distribution for 2022 as bars and overlay them with a prediction line for 2100 that clearly separates from the bars.}'' (P$_{3}$).

\textbf{Comparison to the original chart.}
Ultimately, we compared the individually and collaboratively created charts to the original charts provided by the chart producers. We can see that five of the seven chart producers decided on a line chart for M$_{1}$, which is also the original chart type. While four of these are very similar to the original chart, one participant (P$_{1}$) decided to represent the message through three small line charts, one for each country. The individual charts for M$_{2}$ had many types. While the initial chart was a bar chart, only two participants (P$_{3}$, P$_{7}$) re-designed the message as a stacked bar chart instead of the original grouped bar chart. 

For M$_{3}$, almost all created charts (6 out of 7) were Sankey diagrams, the same chart type as in the original chart. This is unsurprising, as this is a common chart for voter transition analysis, even though they were aware that this chart type is often difficult to understand. To alleviate this problem, interactive Sankey diagrams were discussed.
The last message, M$_{4}$, resulted in five grouped or stacked bar charts, one line chart, and one customized radial plot, while the original chart type was a special bar chart (population pyramid). In contrast to the original population pyramid, the corresponding collaborative chart was a single-sided population pyramid aggregated into age groups.

To summarize, the chart producers were intrigued by creating charts solely based on a given message without having access to the data since they usually start from the data in combination with a message or idea. Interestingly, these created charts were very similar to the original charts overall. However, the producers found it highly beneficial to sketch their charts without thinking about the data and the limitations that might arise from the dataset. While we found that collaborative message creation did not enrich individual messages in the consumer workshops, collaborative chart creation showed great potential as different expert opinions led to essential design considerations in the producer workshop.

Our findings illustrate the dependency of visualization choice on the data, the focus of the message, and the desire to engage audiences, including consideration of implicit associations and emotional effects. This was connected to discussions about reducing mental load.
In observing the producers reworking their individual charts into collaborative ones, we saw how they explained and refined their thinking processes around messages. We could see how they negotiated the intended/desired focus of the message/the chart and how much the visual representation choice depends on the exact intended focus of the message. They realized a diversity of sub-messages were supported by each of their charts.
For instance, ``\textit{In a first step, I assumed that we really only compare 2019 and 2023 point-wise, so only these two [\textit{remark:} data] points.}'' (P-4) illustrates how the presentation and the message shape each other.
Despite the obvious dependence on the data, we saw that also the specifics of the values influence visual representation choices: ``\textit{But with the bars that I created to show the total, [...] you can't really differentiate when there are no big differences -- there is nothing going on}'' (P-4). Answered by another participant with ``\textit{It depends even more on what you said before: about how big the differences [\textit{remark:} in the data] are}'' (P-2).
The consideration of implicit associations is exemplified in ``\textit{I'm not sure about that. I mean, of course, there's a spiritual side to it, where you say, okay, everything's a circle, and we'll all come back one day}'' (P-3).
Increased awareness about those factors that influence practitioners' decision-making processes is needed to better understand takeaway messages for general audiences. The producer workshops showed how the discussions gradually refined the thinking process and sharpened the message.

\textbf{Professional experience.}
Our chart producers have years of professional experience creating and working with charts. We were especially interested in their opinion on different aspects, including aesthetics, trustworthiness, design, and helpfulness of visualizations.

The producers find a chart to have aesthetic appeal when the chart's design elements are well-aligned, meaning that colors, fonts, sizes, space, and forms are carefully chosen and balanced. They should add to a proper understanding of the chart's purpose. Whether the information in a chart should be presented simplified or more complex and nuanced depends primarily on the target audience. A chart is well-designed if the presented information is structured, truthful to the data, and clear. While data visualization is widely used to convey information visually, it is especially helpful for presenting complex or abstract information in a concise way, supporting certain take-away messages, or showing specific characteristics of data (e.g., distributions, trends, changes). We found that chart producers view charts as trustworthy when they contain a (reputable) source and do not distort the data or axes-scales.

\section{Discussion}
\label{sec:discussion}
The results from our study revealed insights that we reflect on and compare to existing literature in this section. Additionally, we give implications for the human-computer interaction and visualization community for communicating data to diverse audiences that can be used to further build on this work and point out the limitations of our work.

\subsection{Reflections}
\textbf{From chart to message and back again: methodological reflections.}
Drawing inspiration from an established method in the evaluation of visualizations~\cite{kerzner2019}, we created a novel approach to study the \textit{gulf of interpretation} between chart producers and diverse (data) visualization consumers. Our method allowed us to examine a diversity of factors influencing chart interpretation, including chart types, subjective factors, and the effects of collaborative sensemaking~\cite{mahyar2014,paul2010} on creating takeaway messages. Reflecting on the success of this method, we find that creating workshops for understanding messages and engaging producers was a rewarding format for both participants and us. 
However, recruiting consumers of various backgrounds is challenging, particularly when reaching senior citizens for such a workshop. Hence, we switched to individual interviews for this group, which is sub-optimal in terms of comparability and provides an open methodological challenge for future research.

The workshop setting was chosen intentionally to create an environment where participants could engage with the visualizations more thoroughly, allowing us to capture detailed insights into their thought processes, interpretation strategies, and challenges. This method provided a level of depth and structure that would be difficult to achieve in less controlled, everyday contexts. While this may differ from typical real-world interactions, it enabled us to observe more detailed behaviors and identify patterns that might remain unnoticed in casual engagements. Future studies could build on these findings by exploring how users engage with visualizations in more naturalistic environments.

Nearly every producer was willing to participate based on the incentive to get detailed feedback on their designs from various slices of the demographic their outlets are trying to reach. They were excited to collaborate with like-minded colleagues in different organizations.
The challenge of improving their designs based on the messages from the consumers was seen as a learning opportunity. For some messages, the graphical convention was too strong (such as Sankey diagrams for changes in voting behaviors), impacting the willingness to explore alternate designs. For less conventional messages, the diversity of designs was striking. The producers relied on their intuition for the design process rather than a structured methodology, as also found by Parsons~\cite{parsons2022}. Starting with the message without showing the detailed data also helped them focus on encoding the message more than the data. Consumers were often overwhelmed with comprehending the data and missing the message. This insight encouraged the producers to abstract the data further and focus on the message in their design process. This means we saw that the focus on the message directly and significantly informs chart design choices.

\textbf{Factors influencing the interpretation of charts.}
We identified several factors that influence the reader's interpretation of charts. High information density in a chart (e.g., due to multiple dimensions [V$_{6}$] or a high number of data points [V$_{7}$]) made it difficult for the participants to articulate a takeaway message because they were overwhelmed. In such cases, participants either focused their message on a specific part of the visualization (a subset of the data) or just repeated the chart's title as they could not decide on a message. While this work specifically focused on static charts, adding interactivity could enhance chart comprehension~\cite{vanberkel2024,pike2009,dimara2020}. However, the potential for exclusion due to the required digital literacy and access to technology must be considered~\cite{aisch2021}. Future work could explore the effects of interactive charts and compare the results.
We observed that the messages are often informed by the text on a chart (e.g., title, annotations). Hence, this text might lead people toward specific parts of the visualization and prime them. On the other hand, missing textual information (e.g., explanations of uncommon terms [V$_{4}$]) resulted in misinterpretations or confusion. This is in line with recent work investigating the role of text on visualizations~\cite{stokes2023,koesten2023}.

In discussing the chart message, participants also referred to its general shape, for instance, ``\textit{The current distribution should look like a Christmas tree, but it looks much more like a kebab}'' (D-2 on V8), or ``\textit{The new population of 2,100 kebabs}'' (D-3 on V8). Additional semantic associations of shape influence the collaborative discussions in particular. Thinking about line charts, there is a known signaling effect of upwards trending curves/lines, which can have emotional connotations as they imply that things are getting better or worse \cite{pokojna2025language}.

The distribution of message types between the consumer groups shows that students' messages mainly (65\%) recalled elemental and encoded properties of the chart (L1). In contrast, more than half of the designers' messages (the most experienced participant group) contained higher levels of semantic content (L3 and L4). 
This could indicate that visualization literacy can be inferred from a consumer's takeaway messages.
If future studies validate connections to literacy levels, messages could be an opportunity to evaluate learning outcomes of data visualization courses, complimentary to standardized tests such as the VLAT~\cite{lee2017}.

Almost 40\% of the participants found charts trustworthy if the chart included a reputable source, which aligns with Peck et al.~\cite{peck2019}, and 28\% mentioned color adding to the chart's aesthetics. In some cases (e.g., Sankey diagram), semantically meaningful color made it easier and faster for them to understand the chart, as also found by Lin et al.~\cite{lin2013} Notably, colors have also shown to influence how well a chart is received across different subjective dimensions if they are purely used as a design element rather than to encode data.

We further observed that familiarity with specific chart types facilitated participants' interpretations of the charts, which aligns with the work by Witt et al.~\cite{witt2021}. However, our study also showed that different populations had difficulties understanding the familiar newspaper charts, which matches findings by Börner et al.~\cite{boerner2016}.

Given our inductive approach, we did not explicitly prompt participants for accessibility issues as this was not the focus of the study. However, especially for the senior citizens, we observed that the size of graphical elements and text is often considered too small, making interpretation of the chart more difficult. (Note that for these instances, the charts were printed in a larger version.) Furthermore, our interviews with retired individuals revealed challenges in interpreting certain charts, which participants attributed to limited data visualization literacy stemming from a lack of formal education in reading diagrams during their schooling.
There is a need for further research to elicit accessibility issues that can and should be used to inform the design of data visualizations for diverse audiences.

\textbf{Individual or collaborative messages -- it does not matter.}
We designed workshops to facilitate individual and collaborative message creation based on studies examining sensemaking in collaborative settings~\cite{mahyar2014,paul2010}. This allowed us to capture rich data and compare individual and collaborative messages. 
To our surprise, when we compared messages by individuals and consolidated by a group of participants, there were not many differences. Overall, collaborative messages were shorter or the same length as individual messages. While this could suggest that creating consolidated messages is not a vital part of our methodology, it could also suggest that collaborative sensemaking needs to be studied by looking at the interactions between participants \textit{in situ} rather than focusing solely on takeaway messages.

\subsection{Implications and Future Work}
Our study indicates the existence of the \textit{gulf of interpretation}. It is our---the visualization community's---responsibility to consider how to bridge this gulf and aid in avoiding misinterpretation or -understanding of visualizations. In the following paragraphs, we provide implications of this work that should encourage our community to question existing conventions. These implications are derived from our study in which we engaged with professional chart producers, diverse chart consumers, and charts that were not artificially created but taken from real-world media. We do not claim generalizability, as this research aimed instead to provide an in-depth qualitative account of the relationship between chart production and interpretation within the specific setting of our study.

\textbf{Rethink conventional chart types.} Our producer interviews showed that some chart types are commonly used to communicate specific data, i.e., seen and used as a standard for particular topics. This is especially true for Sankey diagrams, commonly used to visualize voter shifts in elections, and population pyramids to visualize population distributions of regions. However, the results of our study show that these chart types are often hard to understand. Additionally, our results show that stacked bar charts were especially confusing to our participants, supporting and adding to a variety of research that reports on the low efficacy of stacked bar charts for specific tasks~\cite{cleveland1984,heer2010,talbot2014,indratmo2018}. We urge the visualization community and practitioners to rethink such conventional chart types and develop alternative visual encodings that are more effective and easier to understand. This requires consideration of potential semantic and emotional connotations, such as those associated with colors or shapes used as metaphors. Exploring whether the intentional use of these effects enhances readability and varies across cultural backgrounds presents an interesting direction for future work.

\textbf{Use familiar and basic chart types.} While the chart types mentioned above might be typical for specific topics, our results indicate that familiarity with certain chart types plays an important role in understanding the charts. However, we still know little about what chart types are familiar since different demographic factors might influence this. Our results indicate that in some cases, participants would have preferred a collection of simple charts rather than a single, complex chart that used multiple dimensions. This could be one approach to alleviate the problem with familiarity.

\textbf{Collaborate on visualization design.} The chart producers indicated that time constraints are a big issue that data journalists face when creating visualizations. There is often not enough time to get and incorporate feedback from peers. However, to improve visualization design, our results from the producer workshop indicate that design iterations with multiple visualization experts led to important considerations that individual visualization designers would have missed otherwise. One example is explicit discussions about whether a chart type is not only theoretically suitable for the data but also effective in illustrating a message, considering specific data attributes, values, and the message's visibility. This collaborative chart development facilitates the design process and, hence, might improve the effectiveness of the charts, which can lead to a better understanding for lay audiences.

\textbf{Test the chart.} In addition to the design iterations with multiple visualization experts, our results show that testing the final chart with diverse consumer groups is equally important to improve its efficacy. Since time is often constrained for visualization producers, charts should be evaluated through a lightweight pilot test requiring minimal effort. The discussions in our producer workshops indicated a need for a tool that allows visualization producers to evaluate charts based on pre-defined aspects (e.g., understandability, aesthetics) within a short time frame and which can be easily integrated into their existing work context. Further, such a tool could integrate artificial intelligence features to suggest design improvements while conducting human-in-the-loop evaluations that inform the experts' work practices and lead to better visualization design.

\textbf{Aim for diversity.} While we do not claim representativeness, we included diverse participants who are usually not as prominently featured in visualization research. Conducting workshops and interviews with these groups, especially with young school students and senior citizens, helped us better understand how they make sense of data visualizations. We found discrepancies in the results between young (school and university students) and elderly participants. We must be aware that studies conducted with students (one of the most popular participant groups) do not easily generalize to older people. Our study methodology is an approach to evaluate sensemaking and understanding of data visualizations with diverse audiences. We encourage researchers to adapt and expand the study with a more diverse and potentially representative audience, especially including groups usually overlooked by the visualization community. This inclusivity leads to more generalizable results and insights, making a broader impact possible.

This work is a qualitative contribution to understanding visual data literacy. While some of the implications seem obvious, our main contribution is making our community aware of the gap between intended and received messages, as well as the deficits in the practice of visual data communication in professional mainstream news media. Further, our work shows the complexities of studying messages and their diversity, particularly when considering a diverse audience. While there might be the desire for a single, correct message that people take away from a chart, our results show that messages are multidimensional, complex constructs that we must be aware of.

\subsection{Limitations}
As with every qualitative study that produces in-depth insight, we do not claim representativeness of the population slice we were able to sample. However, it would be interesting to explore the insights of this work further through a large-scale experiment to validate and compare the results, as our results show a significant need to include even more diverse international groups in further studies. Despite our best efforts, we could not unite senior citizens in a workshop setting, making the experimental setting less comparable to the other groups. We also note that some school students faced language proficiency challenges with the term \textit{gender pay gap} in the English title of chart V$_{4}$. However, this concerned only the textual elements on the chart; they were still able to read the data encoded in the stacked bar chart.

While we had a good variety of chart types, many were missing, and the overall sample of charts we investigated in this study was constrained. Expanding the sample to even more chart types could help identify ineffective visualizations. Additionally, since our study focused only on static charts, future research is needed to explore the generalizability of our findings to interactive or animated charts.

\section{Conclusion}
\label{sec:conclusion}
In our work, we confirmed several conventional pearls of wisdom, such that data sources impact trustworthiness~\cite{kosara2019}, proper use of color correlates with aesthetic values~\cite{lim2021}, and bar and line charts tend to be easy to understand~\cite{tversky2011,zacks1999}. However, there were several surprising insights. First, stacked bar charts are harder to understand than Sankey diagrams. Further, consumers are quickly overwhelmed with the amount of data shown, e.g., despite being a \textit{simple} bar chart, an age pyramid can be challenging, likely due to the data details and the rotated axis (a hypothesis that needs to be further validated).

Some visual designs are completely overshadowed by convention, such as Sankey diagrams for change in voting behavior or line charts for (strongly) time-varying data recordings (such as inflation data). Only when explicitly prompted to encode the message (and not the data) are professional chart producers (with many years of experience) able to break from convention and are encouraged to (a) abstract from data towards encoding messages and (b) encouraged to search for better visual encodings.

\begin{acks}
We thank the journalists and our participants for their collaboration, support, and time. We are also very grateful to our colleagues, especially Sebastian Ratzenböck, for their valuable feedback and support. This work has been funded by the Vienna Science and Technology Fund (WWTF) [10.47379/ICT20065].
\end{acks}

\bibliographystyle{ACM-Reference-Format}
\bibliography{refs}

\end{document}